\documentclass{article}
\usepackage{spconf,amsmath,amssymb,graphicx}
\usepackage[hidelinks]{hyperref}
\usepackage{float}
\usepackage{placeins}
\usepackage{fancyhdr}

\title{FLOWSEP-RF: RECTIFIED FLOW MATCHING FOR SINGLE-CHANNEL RADIO-FREQUENCY SIGNAL SEPARATION}
\name{Mohamed Kamel, Tamer Nadeem}
\address{Department of Computer Science, Virginia Commonwealth University, Richmond, VA, USA\\
\{kamelms,tnadeem\}@vcu.edu}
\begin{document}
%
\maketitle
\thispagestyle{fancy}
\begin{abstract}
Single-channel radio-frequency (RF) signal separation is important for interference mitigation when a signal of interest overlaps with a co-channel interferer in time and frequency. Classical matched filtering and LMMSE receivers are limited under structured non-Gaussian interference, while diffusion and score-based methods often require iterative sampling or optimization. We propose FlowSep-RF, a rectified flow matching framework that learns a clean-endpoint transport from the received mixture to the signal of interest. The model combines time-domain IQ samples with frequency-domain features and recovers the separated waveform using one endpoint velocity evaluation. We evaluate FlowSep-RF on QPSK mixtures with OFDM-BPSK, OFDM-QPSK, and RFChallenge CommSignal2 interference against classical, discriminative, and generative baselines. FlowSep-RF achieves the lowest average BER and MSE among evaluated baselines across all three settings with one-step inference.
\end{abstract}


\section{Introduction}
\label{sec:intro}

Co-channel interference is a central challenge in modern wireless systems. In single-channel RF receivers, the signal of interest (SOI) and an interfering waveform may overlap in time and frequency, while spatial filtering is unavailable. The receiver must therefore recover the desired digital signal directly from a mixed baseband waveform.

Classical receivers such as matched filtering and linear minimum mean-square error (LMMSE) filtering are effective when the impairment is well modeled as additive Gaussian noise. However, interference from another communication system is structured and non-Gaussian, so linear filtering may fail to exploit waveform-specific structure. Recent RF source separation benchmarks, including the RF Challenge, have therefore motivated data-driven methods that operate on raw complex IQ samples and evaluate separation using waveform fidelity and downstream demodulation accuracy \cite{Lancho2025RFChallenge,Tse2005FundamentalsWireless}.

Deep neural separators have shown strong performance for RF interference rejection. Prior work exploits temporal structure in cyclostationary signals, while recent U-Net-based and Transformer-based models learn direct mappings from mixtures to separated signals \cite{Lee2022CyclostationaryRF,ronneberger2015unetconvolutionalnetworksbiomedical,lifar2026radiofrequencytransformersignalseparation}. These models provide fast inference, but direct regression does not explicitly model the transport from a corrupted mixture to a clean SOI.

Generative models provide another direction for signal separation. Diffusion and score-based models have been used as learned priors for inverse problems and source separation \cite{3495724.3496298,song2021scorebasedgenerativemodelingstochastic,3524938.3525377}. In RF separation, score-based methods such as randomized Gaussian smoothing exploit the discrete structure of communication signals \cite{jayashankar2024scorebasedsourceseparationapplications}. Related work has also used score-based or diffusion priors for radar interference mitigation and Bayesian signal separation \cite{10889785,11464587}. However, these approaches commonly require iterative sampling, unfolding, or optimization, which can limit their use in low-latency receivers.

Flow matching and rectified flow offer a more efficient generative formulation by learning a velocity field along a path between source and target distributions, avoiding long reverse diffusion chains \cite{Liu2023RectifiedFlow,lipman2023flowmatchinggenerativemodeling,Tong2023ConditionalFM}. Recent signal-processing work has used flow-based models for ECG generation, speech enhancement, and real-time speech restoration \cite{10888856,11462820,11463015,Hsieh2026AdaptiveFlowTSE}.

In this paper, we propose FlowSep-RF, a rectified flow matching framework for single-channel RF signal separation. FlowSep-RF learns a clean-endpoint transport from the received mixture to the SOI, rather than predicting and subtracting the interference. The model combines time-domain IQ modeling with frequency-domain features and recovers the separated waveform using one neural function evaluation.

The main contributions of this work are as follows:
\begin{itemize}
    \item We formulate single-channel RF signal separation as a clean-endpoint rectified flow matching problem from the received mixture to the SOI.
    \item We introduce a dual-domain FlowSep-RF architecture that combines time-domain IQ features with frequency-domain structure for one-step RF separation.
    \item We evaluate FlowSep-RF on QPSK SOI mixtures with OFDM-BPSK, OFDM-QPSK, and RFChallenge CommSignal2 interference, showing improved BER and MSE over classical, discriminative, and generative baselines.
\end{itemize}

\section{Problem Formulation}
\label{sec:problem}

We consider single-channel complex baseband RF signal separation. Let $s \in \mathbb{C}^{N}$ denote the signal of interest (SOI), $b \in \mathbb{C}^{N}$ a co-channel interferer, and $y \in \mathbb{C}^{N}$ the received mixture. The mixture is modeled as
\begin{equation}
    y = s + \kappa b,
    \label{eq:mixture}
\end{equation}
where $\kappa \in \mathbb{R}_{+}$ controls the interference scale. The corresponding signal-to-interference ratio (SIR) is
\begin{equation}
    \gamma_{\mathrm{SIR}} = 10 \log_{10} \frac{\|s\|_{2}^{2}}{\|\kappa b\|_{2}^{2}} .
    \label{eq:sir}
\end{equation}
For unit-power sources, this reduces to $\gamma_{\mathrm{SIR}}=10\log_{10}(1/\kappa^{2})$.

Given $y$, the goal is to recover an estimate $\hat{s}$ of the SOI. Since the true interference scale is unavailable in blind receivers, FlowSep-RF uses a power-based estimate $\hat{\kappa}(y)$, with $\hat{P}_{y}=N^{-1}\sum_{n=1}^{N}|y[n]|^{2}$ and $\hat{\kappa}=\sqrt{\max(0,(\hat{P}_{y}-\hat{P}_{s})/\hat{P}_{b})}$, where $\hat{P}_{s}$ and $\hat{P}_{b}$ are nominal source and interference powers. The corresponding estimated SIR is $\hat{\gamma}_{\mathrm{SIR}}=10\log_{10}(1/\hat{\kappa}^{2})$. During training, the conditioning inputs are jittered as $\tilde{\kappa}=\kappa \xi$, with $\xi\sim\mathcal{U}(0.9,1.1)$, to improve robustness to estimation error. Each complex waveform is represented by its real and imaginary components in $\mathbb{R}^{2\times N}$.

The SOI carries a transmitted bit sequence $\mathbf{u}\in\{0,1\}^{L}$. After separation, the recovered waveform is demodulated as $\hat{\mathbf{u}}=\mathcal{Q}(\hat{s})$. We evaluate waveform fidelity using MSE and communication reliability using BER:
\begin{equation}
\begin{aligned}
    \mathrm{MSE} &= \frac{1}{N}\sum_{n=1}^{N}|\hat{s}[n]-s[n]|^{2}, \\
    \mathrm{BER} &= \frac{1}{L}\sum_{\ell=1}^{L}\mathbf{1}\{\hat{u}_{\ell}\neq u_{\ell}\}.
\end{aligned}
\label{eq:metrics}
\end{equation}
BER is the primary metric because RF separation is ultimately judged by recovery of the transmitted information, while MSE measures waveform reconstruction accuracy.

\section{FlowSep-RF}
\label{sec:method}

\begin{figure*}[t]
    \centering
    \includegraphics[width=0.98\textwidth]{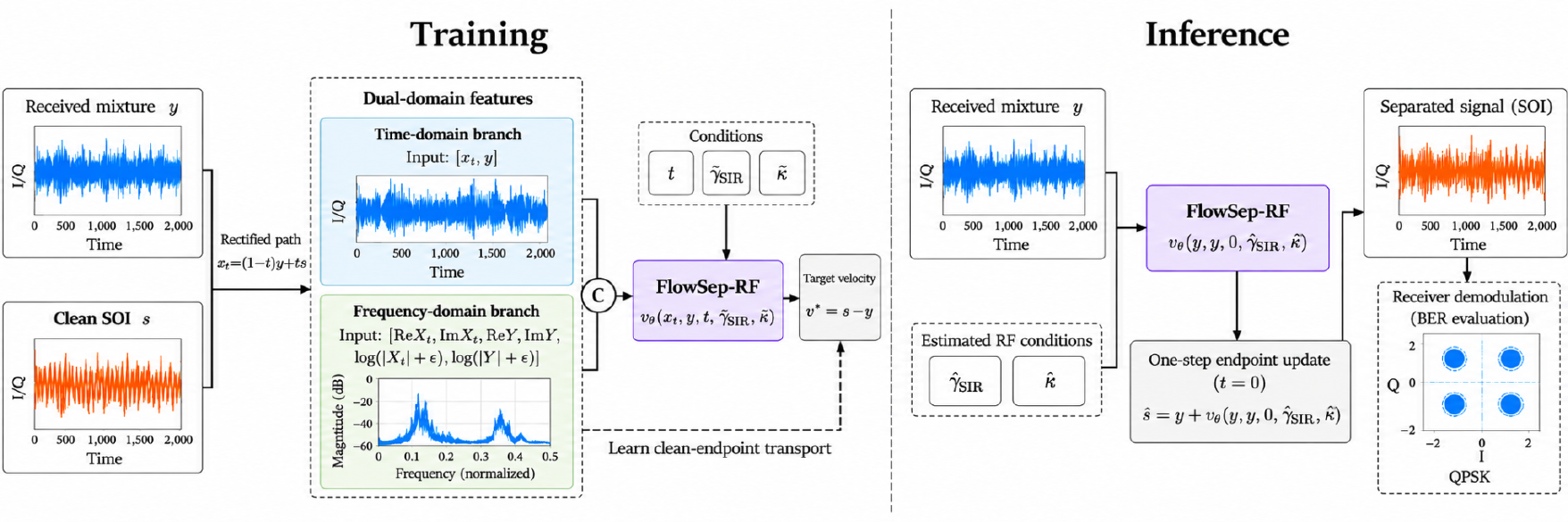}
    \caption{FlowSep-RF overview. Training forms a clean-endpoint path $x_t=(1-t)y+ts$ from the mixture $y$ to the clean SOI $s$ and learns $v^{\star}=s-y$ using time- and frequency-domain features conditioned on RF variables. Inference uses estimated RF conditions and one endpoint update, $\hat{s}=y+v_{\theta}(y,y,0,\hat{\gamma}_{\mathrm{SIR}},\hat{\kappa})$, followed by demodulation for BER evaluation.}
    \label{fig:flowsep_framework}
\end{figure*}

FlowSep-RF separates the SOI by learning a direct transport from the received mixture to the clean waveform, as shown in Fig.~\ref{fig:flowsep_framework}. Unlike interference-prediction methods that estimate $\kappa b$ and subtract it from $y$, FlowSep-RF treats the mixture as the initial endpoint and the SOI as the target endpoint. This gives a clean-endpoint rectified flow formulation aligned with the RF separation objective.

\subsection{Clean-endpoint rectified flow}

For each training example, we set the endpoints to the received mixture and clean SOI and sample $t\sim\mathcal{U}(0,1)$:
\begin{equation}
\begin{aligned}
    x_0 &= y,\qquad x_1=s,\qquad x_t=(1-t)y+ts,\\
    v^{\star} &= x_1-x_0=s-y .
\end{aligned}
\label{eq:rf_path}
\end{equation}
FlowSep-RF learns a conditional velocity field
\begin{equation}
    v_{\theta}=v_{\theta}(x_t,y,t,\tilde{\gamma}_{\mathrm{SIR}},\tilde{\kappa}),
    \label{eq:velocity_field}
\end{equation}
where $\tilde{\kappa}$ denotes the jittered scale parameter during training and the power-based estimate $\hat{\kappa}(y)$ during inference. The mixture $y$ is provided as conditioning information at all times. At inference, the learned field is evaluated at the mixture endpoint, $t=0$, and the separated signal is obtained with one neural function evaluation:
\begin{equation}
    \hat{s}=y+v_{\theta}(y,y,0,\hat{\gamma}_{\mathrm{SIR}},\hat{\kappa}).
    \label{eq:onestep_inference}
\end{equation}
This one-step endpoint update distinguishes FlowSep-RF from iterative diffusion or score-based separators.

\subsection{Dual-domain architecture}

RF mixtures contain useful structure in both time-domain IQ samples and the frequency domain. The time branch receives the real-valued representation of the interpolated sample and mixture,
\begin{equation}
    z_{\mathrm{time}}=[x_t,y]\in\mathbb{R}^{4\times N},
    \label{eq:time_input}
\end{equation}
where the four channels correspond to the real and imaginary components of $x_t$ and $y$. In parallel, the frequency branch applies the discrete Fourier transform to $x_t$ and $y$ and forms
\begin{equation}
    z_{\mathrm{freq}}=[\Re X_t,\Im X_t,\Re Y,\Im Y,\log(|X_t|+\epsilon),\log(|Y|+\epsilon)],
    \label{eq:freq_input}
\end{equation}
where $X_t=\mathcal{F}(x_t)$, $Y=\mathcal{F}(y)$, and $\epsilon$ is a small constant for numerical stability. Both branches are processed by convolutional residual blocks. The time embedding and scalar RF conditions are encoded as $c=\phi(t,\tilde{\gamma}_{\mathrm{SIR}},\tilde{\kappa})$ and injected into the residual blocks through feature-wise modulation. The network predicts a time-domain velocity $v_{\mathrm{time}}$ and a frequency-domain velocity $v_{\mathrm{freq}}$, where the latter is mapped back to the time domain using the inverse Fourier transform. The final velocity is
\begin{equation}
    v_{\theta}=v_{\mathrm{time}}+\alpha v_{\mathrm{freq}},
    \label{eq:dual_velocity}
\end{equation}
where $\alpha$ is a learned scalar gate initialized to give a moderate frequency-domain contribution.

\subsection{Training objective}

The main flow-matching loss trains the model to predict the clean-endpoint velocity along the rectified path:
\begin{equation}
\begin{aligned}
\mathcal{L}_{\mathrm{rf}}=\mathbb{E}_{t}\big[
w(\gamma_{\mathrm{SIR}})
\|v_{\theta}(x_t,y,t,\tilde{\gamma}_{\mathrm{SIR}},\tilde{\kappa})-(s-y)\|_1
\big].
\end{aligned}
\label{eq:rf_loss}
\end{equation}
We also apply an endpoint loss at $t=0$,
\begin{equation}
    \mathcal{L}_{\mathrm{end}}=w(\gamma_{\mathrm{SIR}})\left\|v_{\theta}(y,y,0,\tilde{\gamma}_{\mathrm{SIR}},\tilde{\kappa})-(s-y)\right\|_1,
    \label{eq:end_loss}
\end{equation}
with endpoint reconstruction $\hat{s}_0=y+v_{\theta}(y,y,0,\tilde{\gamma}_{\mathrm{SIR}},\tilde{\kappa})$. To encourage waveform and receiver-level fidelity, we use clean waveform, symbol-proxy, and spectral magnitude losses:
\begin{align}
    \mathcal{L}_{\mathrm{wf}} &= w(\gamma_{\mathrm{SIR}})\|\hat{s}_{0}-s\|_{1},\\
    \mathcal{L}_{\mathrm{sym}} &= w(\gamma_{\mathrm{SIR}})\|\mathcal{P}(\hat{s}_{0})-\mathcal{P}(s)\|_{1},\\
    \mathcal{L}_{\mathrm{spec}} &= w(\gamma_{\mathrm{SIR}})\|\Psi(\hat{s}_{0})-\Psi(s)\|_{1}.
    \label{eq:aux_losses}
\end{align}
Here, $\mathcal{P}(\cdot)$ denotes the receiver-compatible symbol projection used as a differentiable proxy for demodulation quality, and $\Psi(\cdot)$ denotes the log-magnitude Fourier representation. The final training loss is
\begin{equation}
    \mathcal{L}=\lambda_{\mathrm{rf}}\mathcal{L}_{\mathrm{rf}}+\lambda_{\mathrm{end}}\mathcal{L}_{\mathrm{end}}+\lambda_{\mathrm{wf}}\mathcal{L}_{\mathrm{wf}}+\lambda_{\mathrm{sym}}\mathcal{L}_{\mathrm{sym}}+\lambda_{\mathrm{spec}}\mathcal{L}_{\mathrm{spec}} .
    \label{eq:total_loss}
\end{equation}
The SIR weight $w(\gamma_{\mathrm{SIR}})$ up-weights difficult low-SIR and mid-SIR mixtures relative to easier high-SIR examples.

\section{Experimental Setup}
\label{sec:experiments}

\subsection{Datasets}

We evaluate FlowSep-RF on three single-channel RF separation settings. In all experiments, the SOI is an RRC-filtered QPSK waveform and the received mixture is generated using Eq.~\eqref{eq:mixture}. The first two settings use synthetic OFDM interference with BPSK and QPSK modulation, denoted OFDM-BPSK and OFDM-QPSK. The third setting uses real RFChallenge CommSignal2 interference, covering recorded RF interference in addition to synthetic structured interferers.

Each example contains $N=2560$ complex IQ samples, with 160 QPSK symbols and 320 transmitted bits. Mixtures are generated over SIR values of $-24$, $-21$, $-18$, $-15$, $-12$, $-9$, $-6$, and $-3$ dB. For each interference type, we use 200k training mixtures, 20k validation mixtures, and a held-out test grid with 400 examples per SIR level. Test examples are not used during training or model selection.

\subsection{Baselines}

We compare FlowSep-RF with classical, discriminative, and generative baselines. The classical baselines are matched filtering (MF) and LMMSE followed by matched filtering. The discriminative baselines are U-Net-RF and DirectComm-RF, which use one-pass neural inference. The generative baselines are DDPM, reverse diffusion, and $\alpha$-RGS, a score-based RF source separation method \cite{jayashankar2024scorebasedsourceseparationapplications}. These baselines cover linear receivers, fast supervised separators, and iterative generative separation methods.

\begin{figure*}[t]
    \centering
    \begin{minipage}{0.31\textwidth}
        \centering
        \includegraphics[width=\linewidth]{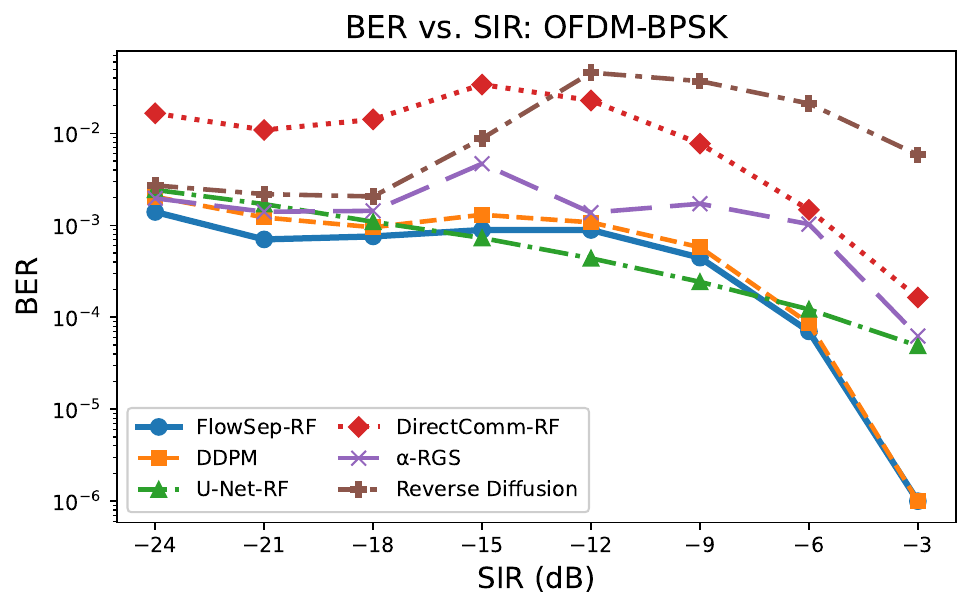}\\
        \footnotesize (a) BER, OFDM-BPSK
    \end{minipage}
    \hfill
    \begin{minipage}{0.31\textwidth}
        \centering
        \includegraphics[width=\linewidth]{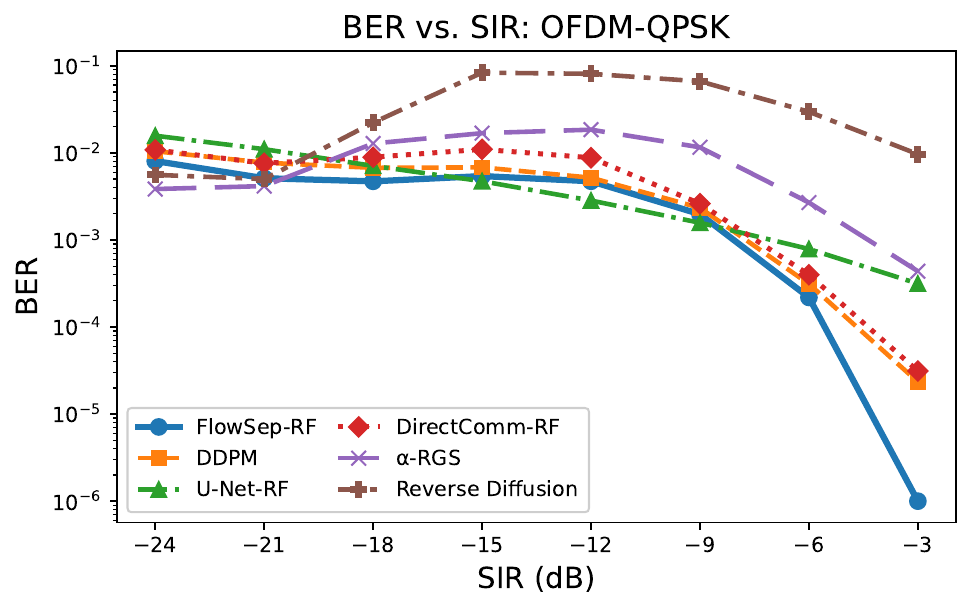}\\
        \footnotesize (b) BER, OFDM-QPSK
    \end{minipage}
    \hfill
    \begin{minipage}{0.31\textwidth}
        \centering
        \includegraphics[width=\linewidth]{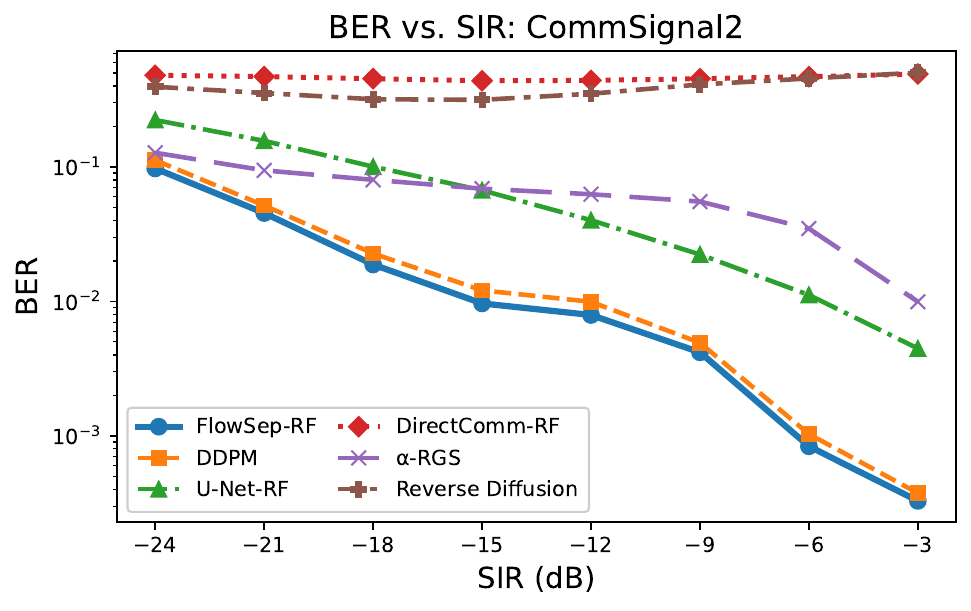}\\
        \footnotesize (c) BER, CommSignal2
    \end{minipage}
    \vspace{0.2em}
    \begin{minipage}{0.31\textwidth}
        \centering
        \includegraphics[width=\linewidth]{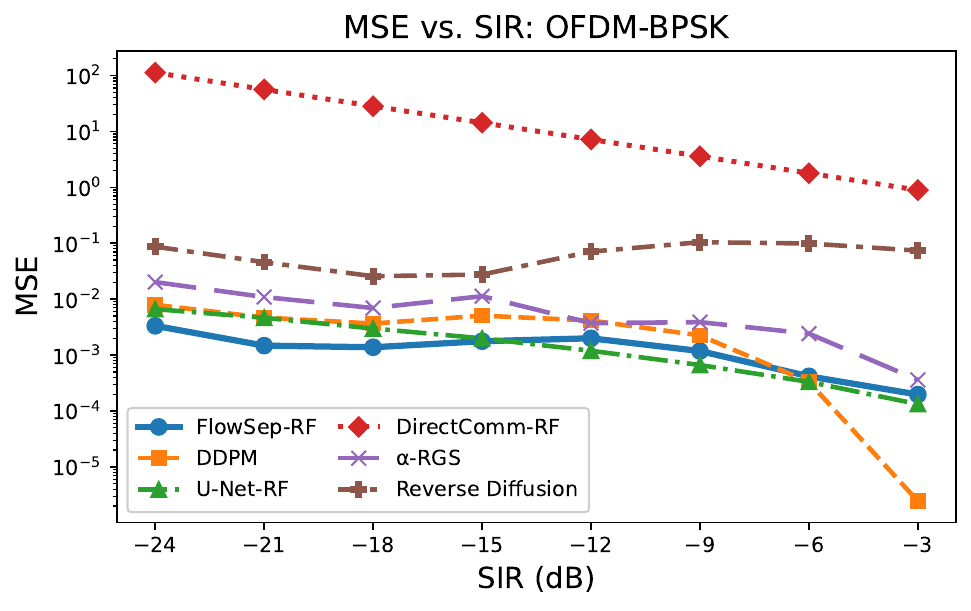}\\
        \footnotesize (d) MSE, OFDM-BPSK
    \end{minipage}
    \hfill
    \begin{minipage}{0.31\textwidth}
        \centering
        \includegraphics[width=\linewidth]{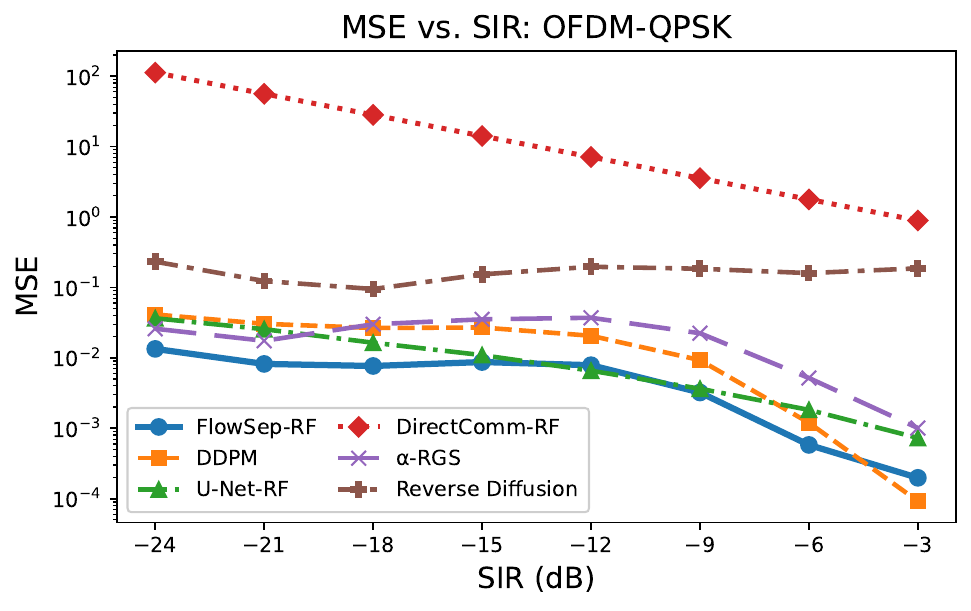}\\
        \footnotesize (e) MSE, OFDM-QPSK
    \end{minipage}
    \hfill
    \begin{minipage}{0.31\textwidth}
        \centering
        \includegraphics[width=\linewidth]{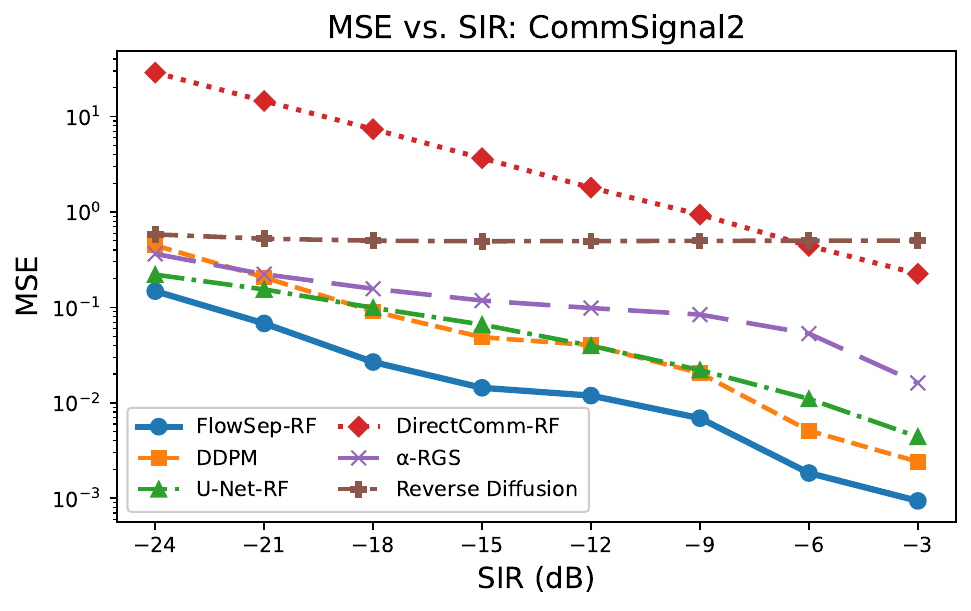}\\
        \footnotesize (f) MSE, CommSignal2
    \end{minipage}
    \caption{BER and MSE versus SIR for FlowSep-RF and baseline methods on the three interference settings. The top row reports BER after demodulation, and the bottom row reports waveform MSE. Lower is better.}
    \label{fig:ber_mse_sir}
\end{figure*}

\subsection{Implementation details}

FlowSep-RF uses the dual-domain architecture described in Section~\ref{sec:method}, with $d_{\mathrm{model}}=192$ hidden channels and eight residual blocks. We train with AdamW using a learning rate of $2\times10^{-4}$, batch size 96, and loss weights $\lambda_{\mathrm{rf}}=1.0$, $\lambda_{\mathrm{end}}=1.0$, $\lambda_{\mathrm{wf}}=0.10$, $\lambda_{\mathrm{sym}}=0.25$, and $\lambda_{\mathrm{spec}}=0.05$. Low-SIR and mid-SIR examples are up-weighted during training. All models are trained for 20 epochs, and the checkpoint with the best validation performance is used for testing.

At inference, FlowSep-RF applies Eq.~\eqref{eq:onestep_inference} with one endpoint velocity evaluation. The conditioning variables use the power-based estimates described in Section~\ref{sec:problem}, while multiplicative jitter during training improves robustness to estimation error. The U-Net-RF and DirectComm-RF baselines use comparable one-pass neural inference, while DDPM, reverse diffusion, and $\alpha$-RGS require iterative sampling or optimization.

We report BER, MSE, and inference efficiency; BER and MSE are defined in Eq.~\eqref{eq:metrics}, while efficiency is measured by test-time neural function evaluations or optimization steps.

\section{Results}
\label{sec:results}

Table~\ref{tab:main_results} reports average BER and MSE over the test SIR grid. FlowSep-RF achieves the lowest average BER and MSE in all three settings using the same configuration. On OFDM-BPSK, it achieves a BER of $6.45\times10^{-4}$ and an MSE of $1.47\times10^{-3}$. On OFDM-QPSK and CommSignal2, it also outperforms DDPM, U-Net-RF, DirectComm-RF, Reverse Diffusion, and $\alpha$-RGS. Fig.~\ref{fig:ber_mse_sir} shows that this trend holds across the SIR grid.

\begin{table}[!t]
\centering
\caption{Main comparison on QPSK SOI separation. BER/MSE are averaged over the test SIR grid. Lower is better; dash indicates unavailable or non-applicable.}
\label{tab:main_results}
\scriptsize
\setlength{\tabcolsep}{3.2pt}
\begin{tabular}{lcccccc}
\hline
\textbf{Method} &
\multicolumn{2}{c}{\textbf{OFDM-BPSK}} &
\multicolumn{2}{c}{\textbf{OFDM-QPSK}} &
\multicolumn{2}{c}{\textbf{CommSignal2}} \\
\cline{2-7}
& \textbf{BER} & \textbf{MSE} & \textbf{BER} & \textbf{MSE} & \textbf{BER} & \textbf{MSE} \\
\hline
MF &
0.186716 & -- &
0.189842 & -- &
0.196826 & -- \\
LMMSE + MF &
0.186716 & 0.904915 &
0.189842 & 0.904157 &
0.196826 & 0.904333 \\
Reverse Diffusion &
0.015707 & 0.066848 &
0.037840 & 0.166426 &
0.387684 & 0.511738 \\
$\alpha$-RGS &
0.001711 & 0.007473 &
0.008855 & 0.021780 &
0.066684 & 0.139131 \\
DirectComm-RF &
0.013484 & 28.138059 &
0.006269 & 28.037159 &
0.027385 & 7.222556 \\
DDPM &
0.000908 & 0.003512 &
0.004946 & 0.019559 &
0.026872 & 0.107607 \\
U-Net-RF &
0.000851 & 0.002324 &
0.005512 & 0.012761 &
0.078125 & 0.077119 \\
\textbf{FlowSep-RF} &
\textbf{0.000645} & \textbf{0.001470} &
\textbf{0.003776} & \textbf{0.006231} &
\textbf{0.023066} & \textbf{0.034845} \\
\hline
\end{tabular}
\end{table}

\noindent\textbf{Inference efficiency.} FlowSep-RF requires one neural function evaluation at the mixture endpoint, matching the one-pass inference cost of U-Net-RF and DirectComm-RF. In contrast, DDPM uses 25 denoising steps, reverse diffusion is iterative, and $\alpha$-RGS uses 2000 test-time optimization steps.

\begin{table}[!h]
\centering
\caption{Ablation study. BER and MSE are averaged over the test SIR grid. Lower is better.}
\label{tab:ablation}
\scriptsize
\setlength{\tabcolsep}{2.5pt}
\resizebox{\columnwidth}{!}{%
\begin{tabular}{lcccccc}
\hline
\textbf{Variant} &
\multicolumn{2}{c}{\textbf{OFDM-BPSK}} &
\multicolumn{2}{c}{\textbf{OFDM-QPSK}} &
\multicolumn{2}{c}{\textbf{CommSignal2}} \\
\cline{2-7}
& \textbf{BER} & \textbf{MSE}
& \textbf{BER} & \textbf{MSE}
& \textbf{BER} & \textbf{MSE} \\
\hline
\textbf{Full FlowSep-RF} &
\textbf{0.000645} & \textbf{0.001470} &
\textbf{0.003776} & \textbf{0.006231} &
\textbf{0.023066} & \textbf{0.034845} \\
w/o rectified flow (direct reg.) &
0.000920 & 0.002150 &
0.005120 & 0.009410 &
0.031200 & 0.045100 \\
w/o symbol-proxy loss &
0.000782 & 0.001615 &
0.004120 & 0.006892 &
0.025140 & 0.038102 \\
w/o spectral loss &
0.000715 & 0.001850 &
0.004050 & 0.007420 &
0.024800 & 0.040150 \\
w/o SIR weighting &
0.000810 & 0.001740 &
0.004310 & 0.007100 &
0.026100 & 0.039800 \\
\hline
\end{tabular}%
}
\end{table}

\noindent\textbf{Ablation study.} Table~\ref{tab:ablation} reports the ablation study. The direct-regression variant uses the same backbone but replaces velocity matching with direct mapping from $y$ to $s$. Removing the rectified-flow objective or any auxiliary component degrades BER and MSE across all three settings. The symbol-proxy, spectral, and SIR-weighting terms improve receiver accuracy, frequency agreement, and low-SIR robustness, respectively.

\section{Conclusion}
\label{sec:conclusion}

We presented FlowSep-RF, a rectified flow matching framework for single-channel RF signal separation. FlowSep-RF learns a clean-endpoint transport from the received mixture to the SOI, combines time- and frequency-domain features, and performs one-step inference. Experiments on QPSK mixtures with synthetic OFDM and RFChallenge CommSignal2 interference show the lowest average BER and MSE among evaluated baselines, and ablations confirm the value of the rectified-flow objective and receiver-aware, spectral, and SIR-aware training terms. These results support clean-endpoint rectified flow for structured co-channel interference.

\section*{Compliance with Ethical Standards}
This is a numerical simulation study for which no ethical approval was required. No funding was received for this study. The authors have no relevant financial or nonfinancial interests to disclose.

\bibliographystyle{IEEEbib}
\bibliography{refs}

\begin{thebibliography}{10}

\bibitem{Lancho2025RFChallenge}
Alejandro Lancho, Amir Weiss, Gary C.~F. Lee, Tejas Jayashankar, Binoy~G. Kurien, Yury Polyanskiy, and Gregory~W. Wornell,
\newblock ``{RF} challenge: The data-driven radio frequency signal separation challenge,''
\newblock {\em IEEE Open Journal of the Communications Society}, vol. 6, pp. 4083--4100, 2025.

\bibitem{Tse2005FundamentalsWireless}
David Tse and Pramod Viswanath,
\newblock {\em Fundamentals of Wireless Communication},
\newblock Cambridge University Press, 2005.

\bibitem{Lee2022CyclostationaryRF}
Gary C.~F. Lee, Amir Weiss, Alejandro Lancho, Jennifer Tang, Yuheng Bu, Yury Polyanskiy, and Gregory~W. Wornell,
\newblock ``Exploiting temporal structures of cyclostationary signals for data-driven single-channel source separation,''
\newblock in {\em 2022 IEEE International Workshop on Machine Learning for Signal Processing (MLSP)}, 2022, pp. 1--6.

\bibitem{ronneberger2015unetconvolutionalnetworksbiomedical}
Olaf Ronneberger, Philipp Fischer, and Thomas Brox,
\newblock ``{U-Net}: Convolutional networks for biomedical image segmentation,'' 2015.

\bibitem{lifar2026radiofrequencytransformersignalseparation}
Egor Lifar, Semyon Savkin, Rachana Madhukara, Tejas Jayashankar, Yury Polyanskiy, and Gregory~W. Wornell,
\newblock ``The radio-frequency transformer for signal separation,'' 2026.

\bibitem{3495724.3496298}
Jonathan Ho, Ajay Jain, and Pieter Abbeel,
\newblock ``Denoising diffusion probabilistic models,''
\newblock in {\em Proceedings of the 34th International Conference on Neural Information Processing Systems}, Red Hook, NY, USA, 2020, NIPS '20, Curran Associates Inc.

\bibitem{song2021scorebasedgenerativemodelingstochastic}
Yang Song, Jascha Sohl-Dickstein, Diederik~P. Kingma, Abhishek Kumar, Stefano Ermon, and Ben Poole,
\newblock ``Score-based generative modeling through stochastic differential equations,'' 2021.

\bibitem{3524938.3525377}
Vivek Jayaram and John Thickstun,
\newblock ``Source separation with deep generative priors,''
\newblock in {\em Proceedings of the 37th International Conference on Machine Learning}. 2020, ICML'20, JMLR.org.

\bibitem{jayashankar2024scorebasedsourceseparationapplications}
Tejas Jayashankar, Gary C.~F. Lee, Alejandro Lancho, Amir Weiss, Yury Polyanskiy, and Gregory~W. Wornell,
\newblock ``Score-based source separation with applications to digital communication signals,'' 2024.

\bibitem{10889785}
Xinyi Wei, Jihwan Youn, Jeroen Overdevest, Jun Li, Satish Ravindran, and Ruud~J.G. van Sloun,
\newblock ``Deep unfolding using score-based generative networks for automotive radar interference mitigation,''
\newblock in {\em ICASSP 2025 - 2025 IEEE International Conference on Acoustics, Speech and Signal Processing (ICASSP)}, 2025, pp. 1--5.

\bibitem{11464587}
Yi~Zhang, Rui Guo, and Yonina~C. Eldar,
\newblock ``Bayesian signal separation via plug-and-play diffusion-within-gibbs sampling,''
\newblock in {\em ICASSP 2026 - 2026 IEEE International Conference on Acoustics, Speech and Signal Processing (ICASSP)}, 2026, pp. 16--20.

\bibitem{Liu2023RectifiedFlow}
Xingchao Liu, Chengyue Gong, and Qiang Liu,
\newblock ``Flow straight and fast: Learning to generate and transfer data with rectified flow,''
\newblock in {\em International Conference on Learning Representations}, 2023.

\bibitem{lipman2023flowmatchinggenerativemodeling}
Yaron Lipman, Ricky T.~Q. Chen, Heli Ben-Hamu, Maximilian Nickel, and Matt Le,
\newblock ``Flow matching for generative modeling,'' 2023.

\bibitem{Tong2023ConditionalFM}
Alexander Tong, Kilian Fatras, Nikolay Malkin, Guillaume Huguet, Yanlei Zhang, Jarrid Rector-Brooks, Guy Wolf, and Yoshua Bengio,
\newblock ``Conditional flow matching: Simulation-free dynamic optimal transport,'' 2023.

\bibitem{10888856}
Yuta Nambu, Masahiro Kohjima, and Ryuji Yamamoto,
\newblock ``Cardioflow: Learning to generate ecg from ppg with rectified flow,''
\newblock in {\em ICASSP 2025 - 2025 IEEE International Conference on Acoustics, Speech and Signal Processing (ICASSP)}, 2025, pp. 1--5.

\bibitem{11462820}
Naisong Zhou, Saisamarth~Rajesh Phaye, Milos Cernak, Tijana Stojković, Andy Pearce, Andrea Cavallaro, and Andy Harper,
\newblock ``Shortcut flow matching for speech enhancement: Step-invariant flows via single stage training,''
\newblock in {\em ICASSP 2026 - 2026 IEEE International Conference on Acoustics, Speech and Signal Processing (ICASSP)}, 2026, pp. 15842--15846.

\bibitem{11463015}
Tsun-An Hsieh and Sebastian Braun,
\newblock ``Towards real-time generative speech restoration with flow-matching,''
\newblock in {\em ICASSP 2026 - 2026 IEEE International Conference on Acoustics, Speech and Signal Processing (ICASSP)}, 2026, pp. 15847--15851.

\bibitem{Hsieh2026AdaptiveFlowTSE}
Tsun-An Hsieh and Minje Kim,
\newblock ``Adaptive deterministic flow matching for target speaker extraction,''
\newblock in {\em ICASSP 2026 - 2026 IEEE International Conference on Acoustics, Speech and Signal Processing (ICASSP)}, 2026, pp. 15412--15416.

\end{thebibliography}

\end{document}